\documentclass[aps,reprint, prb, longbibliography, superscriptaddress, showpacs,nobibnotes,floatfix]{revtex4-1}
\usepackage{graphicx}
\usepackage{bm,amsmath}
\usepackage{dcolumn}
\usepackage{natbib,xcolor}
\usepackage[version=4]{mhchem}
\usepackage{upgreek}
\usepackage{amsmath}

\begin{document}
\title{Anisotropic magnetotransport properties coupled with spiral spin modulation in a triangular-lattice magnet EuZnGe}

\author{Takashi Kurumaji}
\affiliation{Department of Advanced Materials Science, University of Tokyo, Kashiwa 277-8561, Japan}
\author{Masaki Gen}
\affiliation{Department of Advanced Materials Science, University of Tokyo, Kashiwa 277-8561, Japan}
\affiliation{RIKEN Center for Emergent Matter Science (CEMS), Wako 351-0198, Japan}
\author{Shunsuke Kitou}
\affiliation{RIKEN Center for Emergent Matter Science (CEMS), Wako 351-0198, Japan}
\author{Hajime Sagayama}
\affiliation{Institute of Materials Structure Science, High Energy Accelerator Research Organization, 305-0801 Tsukuba, Japan}
\author{Akihiko Ikeda}
\affiliation{Department of Engineering Science, University of Electro-Communications, Chofu, Tokyo 182-8585, Japan}
\author{Taka-hisa Arima}
\affiliation{Department of Advanced Materials Science, University of Tokyo, Kashiwa 277-8561, Japan}
\affiliation{RIKEN Center for Emergent Matter Science (CEMS), Wako 351-0198, Japan}
\date{\today}
\begin{abstract}
We investigate the thermodynamic, magnetic, and electrical transport properties of a triangular-lattice antiferromagnet EuZnGe using single crystals grown from Eu-Zn flux in sealed tantalum tubes.
Magnetic properties are found to be isotropic in the paramagnetic state while we observe an enhancement of in-plane magnetic susceptibility at the temperature near $T^{*}=$11.3 K, suggesting an easy-plane anisotropy at low temperatures.
Magnetic transition temperature is lower than $T^{*}$ as specific heat shows a peak at $T_{\text{N}}=$7.6 K.
We reveal the magnetic modulation along the $c$ axis by resonant x-ray scattering at Eu $L_2$ edge, which suggests competing magnetic interaction among Eu triangular-lattice layers.
We observe a double-peak structure in the intensity profile along (0, 0, $L$) below $T_{\text{N}}$, which is mainly composed of a dominant helical modulation with $\bm{q}\sim$ (0, 0, 0.4) coexisting with a secondary contribution from $\bm{q}\sim$ (0, 0, 0.5).
We reproduce the intensity profile with a random mixture of five- and four-sublattice helices with spin rotation skipping due to hexagonal in-plane anisotropy.
The metallic conductivity is highly anisotropic with the ratio $\rho _{zz}/\rho_{xx}$ exceeding 10 over the entire temperature range and additionally exhibits a sharp enhancement of $\rho_{zz}$ at $T_{\text{N}}$ giving rise to  $\rho _{zz}/\rho_{xx}\sim 50$, suggesting a coupling between out-of-plane electron conduction and the spiral magnetic modulations.
In-plane magnetic field induces a spin-flop like transition, where the $q = 0.4$ peak disappears and an incommensurate peak of approximately $q_{\text{ICM}}\sim$ 0.47 emerges, while the $q=0.5$ modulation retains a finite intensity.
This transition correlates with non-monotonic magnetoresistance and Hall resistivity, suggesting a significant interplay between electrons and spin structures through Ruderman-Kittel-Kasuya-Yosida (RKKY) interaction.
\end{abstract}

\maketitle

\section{Introduction}
Interplay between conduction electrons and underlying magnetic textures is one of the central interests in modern condensed matter physics.
Discoveries of new quantum materials showing novel magnetotransport responses are essential to facilitate future application to spintronic devices.
Recent advances of this research field are, for example, discoveries of quantum anomalous Hall effect in magnetic topological insulators \cite{chang2013experimental,deng2020quantum}, spontaneous anomalous Hall effect and Nernst effect in magnetic Weyl metals \cite{nakatsuji2015large, liu2018giant}, and topological Hall effect (THE) associated with non-coplanar spin textures such as skyrmion lattice states \cite{neubauer2009topological,muhlbauer2009skyrmion,kurumaji2019skyrmion}.

Europium-based intermetallics are a group of materials having been investigated for long time and attracting recent interests as a source of various topological spin textures \cite{kakihana2018giant,shang2021anomalous,takagi2022square,moya2021incommensurate}, and exotic transport phenomena including anomalous Hall effect associated with topological band structures \cite{ma2019spin,xu2019higher,soh2019ideal,gui2019new,xu2021unconventional,pierantozzi2022evidence}, giant magnetoresistance \cite{meul1982transport,sullow1998structure,yin2020large,rosa2020colossal}, coexistence of superconductivity and helimagnetism \cite{iida2019coexisting}, three-dimensional quantum Hall effect \cite{masuda2016quantum}, and valence transitions \cite{onuki2020unique}.

EuZnGe, the material of interest in this study, crystallizes in the centrosymmetric ZrBeSi type ($P6_3/mmc$) structure (Fig. 1(a)), which belongs to one of the $RMX$ phases ($R$: alkali earths/europium/ytterbium, $M$: Cu-Au/Zn-Cd, $X$: anionic main group elements) with \ce{AlB2} as the aristotype structure \cite{Merlo,pottgen2000equiatomic}.
Every other layer of the Zn-Ge honeycomb network is rotated by 60$^{\circ}$ with respect to each other, and they are stacked along the $c$ axis sandwiching triangular lattices of Eu.
Formal valence of each atom is Eu$^{2+}$Zn$^{2+}$Ge$^{4-}$, which can be described as an electron-precise Zintl phase \cite{matar2019coloring}.
Finite density of state at $E_{\text{F}}$ is ascribed to the overlap between the conduction and valence bands, resulting in a metallic behavior \cite{pottgen2000equiatomic}.
Previous studies of magnetism \cite{Merlo,EuZnGe,pottgen2000equiatomic} were performed with polycrystalline samples, reporting divalent Eu$^{2+}$ nature with localized magnetic moments.
Two antiferromagnetic-like transitions at 12.2 K and 9.5 K were observed though the dominant magnetic interaction was suggested to be ferromagnetic by the Weiss temperature $\mathit{\Theta_{\text{W}}=}$ 11 K.
Together with a signature of metamagnetic transition at 0.6 T before the saturation, magnetic properties suggest the presence of inherent magnetic frustration in Eu-sublattice in EuZnGe.

In this work, we report the nature of magnetism and electric properties of single crystals of EuZnGe.
We combine transport and thermodynamic measurements with resonant x-ray scattering (RXS) to reveal a strong correlation between magnetotransport properties and the spin structures with a double-modulation peak along the $c$ axis.
We simulate the spin structure to reproduce the magnetic diffraction profile to propose a helical configuration that is modified by the spin-skipping due to hexagonal in-plane anisotropy.

\section{Experimental methods}
Single crystals of EuZnGe were grown from Eu-Zn flux in a sealed tantalum crucible.
The method proposed in Ref. \onlinecite{volatile_flux} was applied to avoid the reaction of Eu with quartz tube and prevent the loss of volatile Eu-Zn flux.
As shown in the ternary phase diagram (Fig. 1(b)), synthesis of the EuZnGe phase in Zn-rich flux (m.p.: 420$^{\circ}$C) is inhibited by the tetragonal phase \ce{EuZn2Ge2} \cite{Grytsiv} (green line).
We focused on the Eu-Zn flux as the binary phase diagram (Fig. 1(c)) exhibits low-temperature liquidus line below 800$^{\circ}$C.
The europium ingots (99.9\%), zinc wires (99.99\%), and germanium pieces (99.999\%) were put into a tantalum tube with the molar ratio Eu:Zn:Ge = 2:2:1 (the composition for the flux 3, red cross in Fig. 1(b)), which was sealed by an arc furnace under Ar atmosphere.
Eu ingots were handled in an Ar-filled glovebox.
The tantalum tube was put into an evacuated quartz tube and heated to 950$^{\circ}$C.
After being held for 24 hours, it was then cooled to 750$^{\circ}$C in 200 hours.
After annealing for 4 days at this temperature, the excess flux was removed by centrifugation.
\begin{figure}[t]
	\includegraphics[width = \columnwidth]{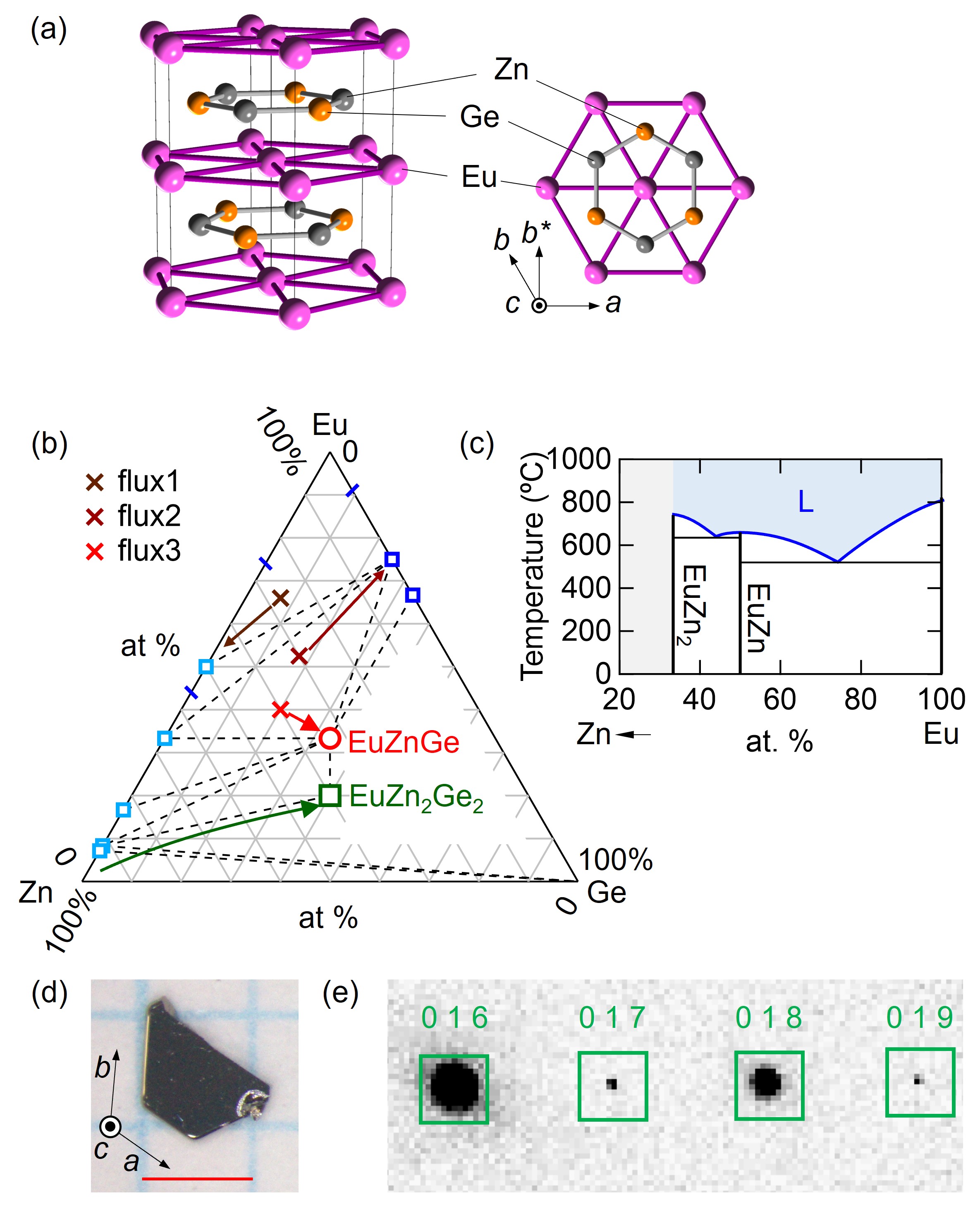}
	\caption{\label{fig1} (a) Crystal structure of EuZnGe.
	Top view is on the right, where $b^{*}$ axis is defined as the in-plane axis perpendicular to the $a$ axis.
	(b) Schematic convex hull diagram for ternary Eu-Zn-Ge, reconstructed using the Materials Projects (materialsproject.org) and the Open Quantum Materials Database (oqmd.org).
	Blue bars are eutectic points, open squares and a circle are equilibrium phases.
	Crosses are starting composition for flux growths.
	(c) Binary phase diagram of Eu-Zn, reproduced from Ref. \onlinecite{ASM}.
	(d) A picture of a single crystal of EuZnGe.
	Red scale bar represents 1 mm.
	(e) Bragg peaks of EuZnGe observed in the single-crystal XRD using the synchrotron light source at room temperature.
}
\end{figure}

The atomic composition and the phase were checked by energy dispersive x-ray spectroscopy (EDS, JEOL model JSM-6010LA) and a Rigaku SmartLab diffractometer using Cu $K_{\alpha}$ radiation, respectively.
The crystal structure was investigated by a single-crystal x-ray diffractometer at the synchrotron facility SPring-8 (see Appendix).
A He-gas-blowing device was employed to cool the crystal to 50 K.

Magnetization was measured with a superconducting quantum interference device magnetometer (Quantum Design MPMS-XL).
Electrical transport measurements were performed by a conventional five probe method at typical frequency near 17 Hz.
The transport properties at low temperatures in a magnetic field was measured using a commercial superconducting magnet and cryostat.
The obtained longitudinal and transverse resistivities were symmetrized and antisymmetrized, respectively by field to correct contact misalignment.
Thermal expansion and the magnetostriction was measured by the fiber-Bragg-grating (FBG) technique using an optical sensing instrument (Hyperion si155, LUNA) in an Oxford Spectromag as described in Ref. \onlinecite{gen2022enhancement}.
Optical fibers were glued using epoxy (Stycast1266) on the (001) and (100) surfaces of as-grown crystals to measure the elongation/compression along the $a$ and $c$ axes, respectively.

Single-crystal RXS measurement was carried out at BL-3A, Photon Factory, KEK, Japan, by using the horizontally polarized x-ray in resonance with Eu $L_{2}$ absorption edge (7.615 keV).
We attached a crystal with the as-grown (001) plane on an aluminum plate with GE varnish, and loaded it into a vertical-field superconducting magnet with the $b$ axis parallel to the magnetic field direction.
The scattering plane was set to be ($H$, 0, $L$).
Unless stated, scattered x-rays were detected without polarization analysis.

\section{Results}
We obtained single crystals of thin hexagonal-prism shape with the typical dimension 1$\times$1$\times$0.5 mm$^{3}$, as shown in Fig. 1(d).
The atomic composition was checked to be Eu:Zn:Ge = 0.35:0.33:0.32, which fairly agrees with the nominal chemical formula.
No trace of tantalum was detected in the crystals.
Powder x-ray diffraction (XRD) patterns confirmed the single phase of EuZnGe.
In the single-crystal XRD with synchrotron radiation, we observed odd $l=2n+1$ reflections as shown in Fig. 1(e), exemplifying a doubled-cell of the ZrBeSi-type structure along the $c$ axis with respect to the \ce{AlB2}-type unit cell due to the Zn/Ge-site ordering as shown in Fig. 1(a) (see Appendix).
We refined the crystal structure and discerned negligible site mixing for Zn/Ge atoms or substitution of Ge for Zn as reported in Ref. \onlinecite{EuZnGeZintl}.
The refined lattice constants at room temperature are $a =$ 4.3700(5) \AA$ $ and $c$ = 8.5994(5) \AA, which is consistent with the previously reported values \cite{EuZnGe} in polycrystalline samples.
We also succeeded in synthesizing single crystals of BaZnGe with the same method. 
The lattice constants are $a=$4.47 \AA $ $ and $c=$9.62 \AA, consistent with that reported in Ref. \onlinecite{Merlo}.
Single crystals of EuZnGe are stable in air while those of BaZnGe are oxidized within a few days.
Prior to the growth with flux 3, we attempted different growth paths starting from flux 1 and 2 (brown crosses in Fig. 1(b)), which are on the line connecting EuZnGe and Eu$_{0.74}$Zn$_{0.26}$, the lowest Eu-Zn eutectic point.
We failed obtaining EuZnGe phase in these runs.
EuZn or Eu$_{3}$Ge appeared instead as the dominant phases, potentially because of their compositional proximity and/or high melting point.

\begin{figure}[t]
	\includegraphics[width =  \columnwidth]{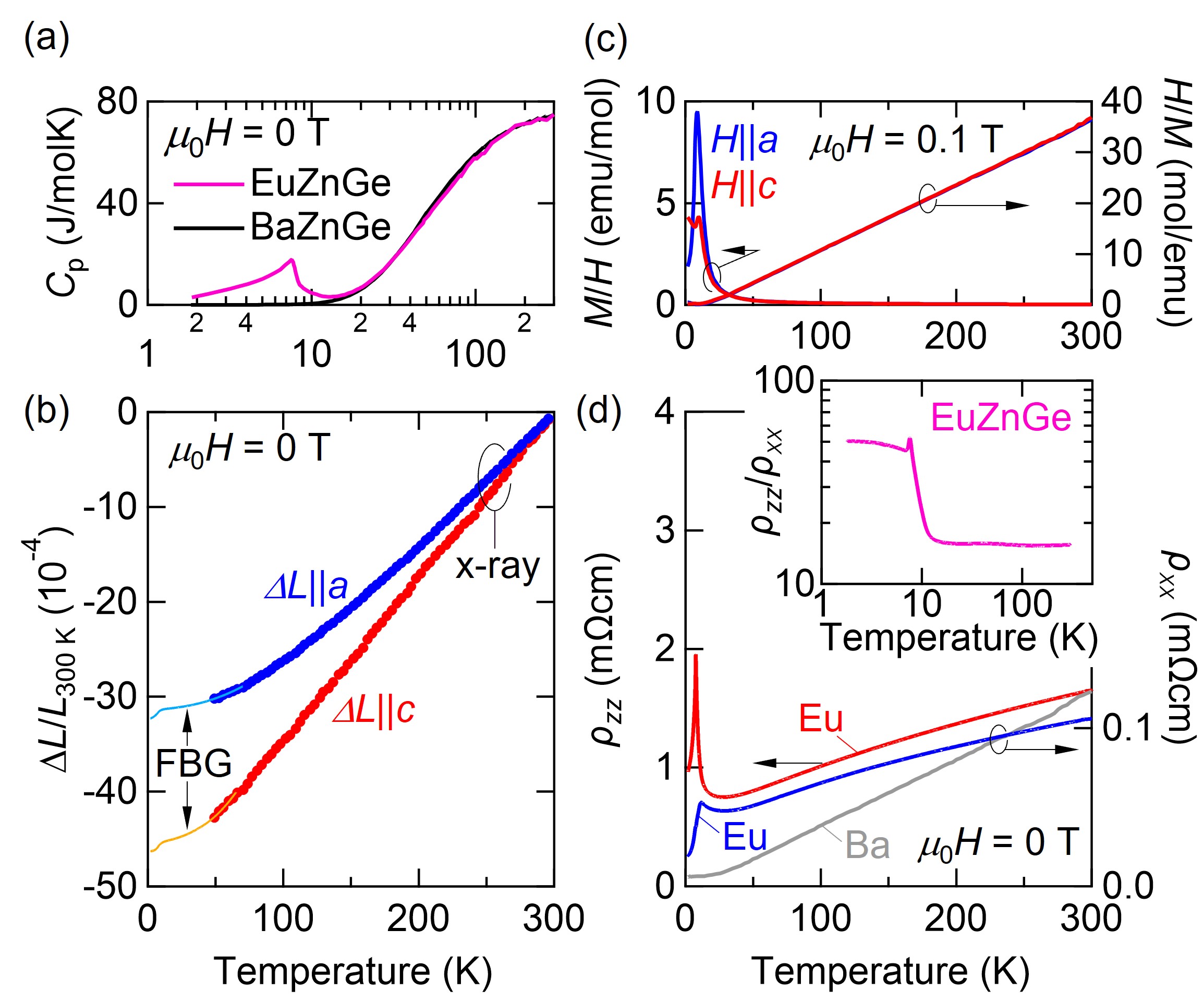}
	\caption{\label{fig2} Temperature dependence of physical properties of EuZnGe.
	(a) Specific heat ($C_{\text{p}}$) in zero field, (b) thermal expansion ($\Delta L/L_0=(L(T,\mu _0H)-L_0)/L_0$) measured by the single-crystal XRD (closed circle) and FBG (thin line), (c) magnetic susceptibility ($M/H$) and inverse susceptibility ($H/M$), and (d) out-of-plane ($\rho_{zz}$) and in-plane ($\rho_{xx}$) resistivity.
	Inset of (d) is the resistivity ratio $\rho_{zz}/\rho_{xx}$ of EuZnGe.
	In (a) and (d), data for BaZnGe is also shown.}
\end{figure}

We summarize the temperature dependence of physical properties of single crystalline EuZnGe in Fig. 2.
The magnetic transition at $T_{\text{N}}=$ 7.6 K can be detected as the specific heat peak at zero field (Fig. 2(a)).
The high-temperature feature is dominated by phonons with no signature of structural phase transition as is similar with that of isostructural BaZnGe.
The absence of structural transition is also confirmed by single-crystal XRD and thermal expansion at zero field (Fig. 2(b)).
The magnetic susceptibilities ($M/H$) for $H\parallel c$ and $H\parallel a$ (Fig. 2(c)) exhibit Curie-Weiss feature characteristic for localized spins of Eu$^{2+}$ in the paramagnetic state (Fig. 2(c)).
The inverse susceptibility ($H/M$) gives the estimation of effective moments as $p_{\text{eff}\parallel c}=$ 7.89 $\mu_{\text{B}}$ for $H\parallel c$ and $p_{\text{eff}\perp c}=$ 7.92 $\mu_{\text{B}}$ for $H\parallel a$, both of which are close to 7.94 $\mu_{\text{B}}$ for free Eu$^{2+}$ ions with $S=\frac{7}{2}$.
Weiss temperatures are $\mathit{\Theta_{\text{W}\parallel c}}=$ 16.2 K, and $\mathit{\Theta_{\text{W}\perp c}=}$ 15.7 K for $H\parallel c$ and $H\perp c$, respectively, suggesting ferromagnetic interaction as the dominant coupling.
At low temperatures, easy-plane type anisotropy starts to develop just above $T_{\text{N}}$.
Figure 2(d) compares resistivity for in-plane and out-of-plane current directions.
The resistivity ratio $\rho _{zz}/\rho _{xx}$ (see inset) exceeds ten at all temperatures, consistent with the layered crystal structure with quasi-2D nature of Fermi surface.
At $T_{\text{N}}$, $\rho _{zz}/\rho _{xx}$ reaches near fifty, which we attribute to enhanced spin-electron scattering for out-of-plane transport.

\begin{figure}[t]
	\includegraphics[width =  \columnwidth]{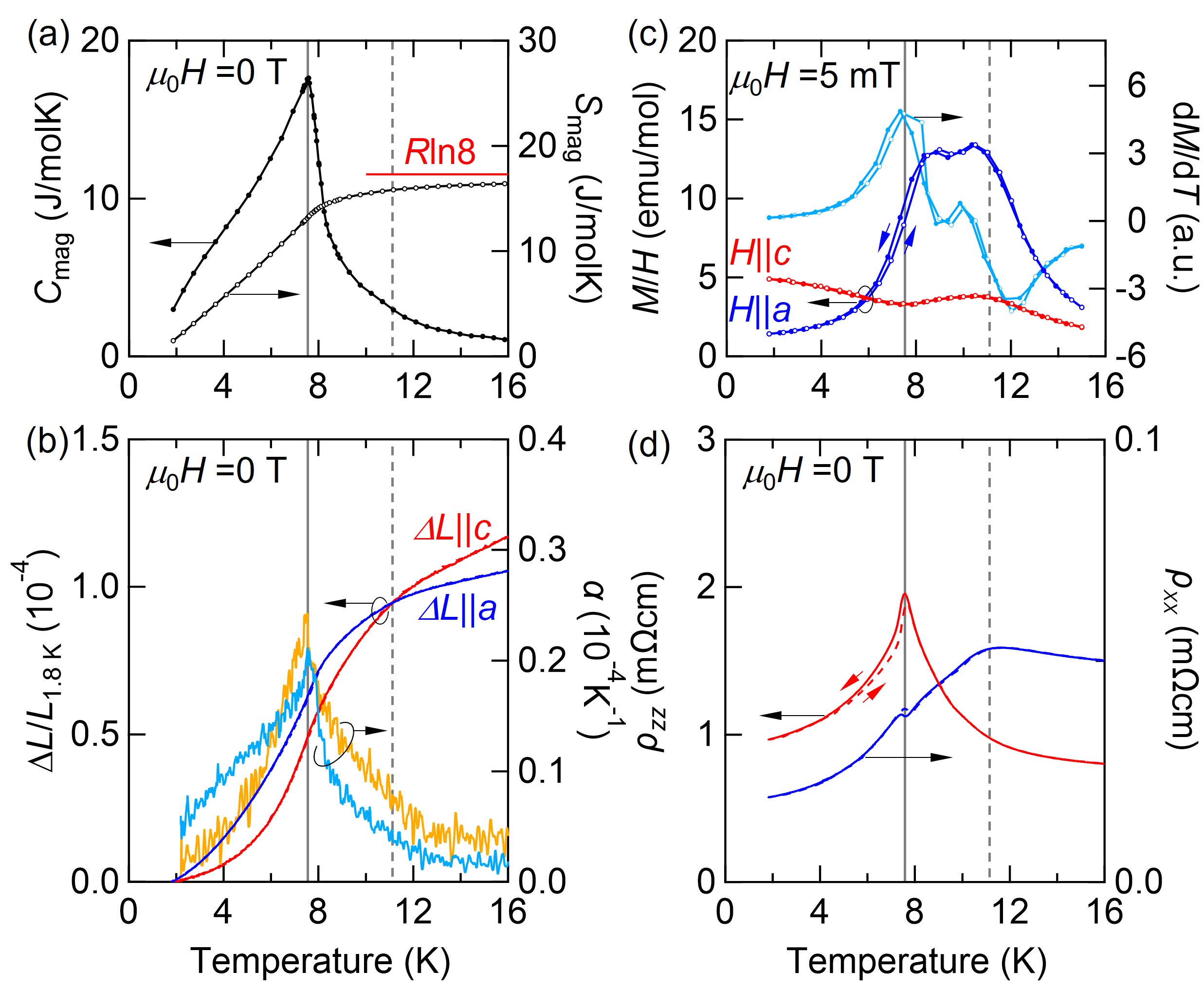}
	\caption{\label{fig3} Temperature dependence of physical properties of EuZnGe.
	Vertical solid gray lines denote the transition temperature $T_{\text{N}}$, and the dashed lines are for $T^{*}=$11.3 K (see text).
	(a) Magnetic specific heat ($C_{\text{mag}}$) and magnetic entropy ($S_{\text{mag}}$) in zero field, (b) thermal expansions ($\Delta L/L_0$) and thermal-expansion coefficient $\alpha$ ($=\text{d}(\Delta L/L_0)/\text{d}T$) along the $c$ (red/orange curve) and $a$ (blue/cyan curve) axes. 	In (a), $S_{\text{mag}}$ at $T=1.8$ K is shifted by $S_{\text{mag0}}=C_{\text{mag, 1.8 K}}/2$.
	(c) Magnetic susceptibility ($M/H$) for $H\parallel a$ (blue curve) and $H\parallel c$ (red curve).
	The temperature derivative of $M$ ($\text{d}M/\text{d}T$) for $H\parallel a$ (cyan curve) is also shown.
	Closed (open) symbols are for cooling (warming) process.
	(d) Out-of-plane (red, $\rho_{zz}$) and in-plane (blue, $\rho_{xx}$) resistivity.
	Solid (dashed) curve is for cooling (warming) process.
	}
\end{figure}

Next we move on to the low-temperature behavior associated with the magnetic ordering.
We summarize the physical properties near $T_{\text{N}}$ in Fig. 3.
A clear single peak in $C_{\text{mag}}$ was observed at $T_{\text{N}}$.
The magnetic entropy $S_{\text{mag}}$ is obtained by the integration of $C_{\text{mag}}/T$ with respect to $T$ (Fig. 2(a)), and approaches $R\ln (2S+1)$ with $S=\frac{7}{2}$ at well above $T_{\text{N}}$.
This is consistent with the idea that the Eu$^{2+}$ ions dominate the magnetism.
The onset of the magnetic order at $T_{\text{N}}$ coincides with a steep change in the lattice distortion for both along $a$ and $c$ axes (Fig. 3(b)), which is clear in the thermal expansion coefficient $\alpha$ ($=\text{d}(\Delta L/L)/\text{d}T$).
The profiles of $C_{\text{mag}}$ and $\alpha$ reasonably agree with each other for the second order phase transition, indicating that the spin-lattice coupling does not appear to significantly facilitate the magnetic transition at $T_{\text{N}}$ \cite{testardi1975elastic}.

\begin{figure}[t]
	\includegraphics[width =  \columnwidth]{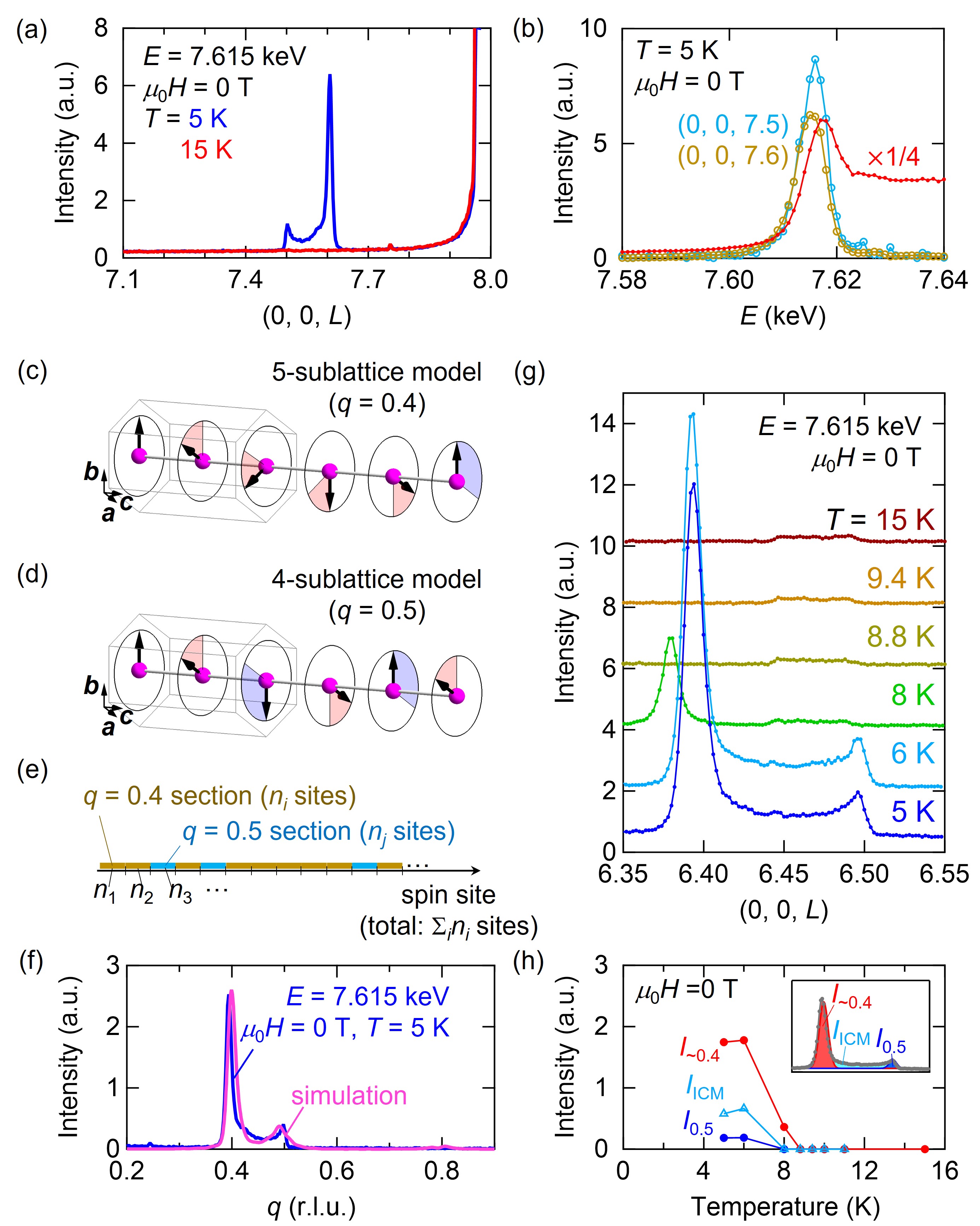}
	\caption{\label{fig4}
	(a) Intensity profile in the (0, 0, $L$) scan with the incident energy $E =$ 7.615 keV in zero field at $T =$ 5 K (blue), and 15 K (red).
	(b) Energy scans of magnetic Bragg spots at (0, 0, 7.5) and (0, 0, 7.4) (cyan and yellow symbols, respectively), and fluorescence (red curve).
	(c)-(d) Schematic possible Eu-spin structures: (c) five-sublattice spiral structure for $q=$0.4; and (d) four-sublattice spiral structure for $q=$0.5.
	Spin rotation across neighboring layers is hypothesized to be fixed at either 60$^{\circ}$ or 120$^{\circ}$ as denoted by pink and blue fan, respectively.
	(e) Schematic spin configuration for the randomized spin-skipping model (see text) in one-dimensional spin chain.
	Yellow and cyan segments are $q=$ 0.4 and $q=$0.5 sections of $n_i$ and $n_j$ sites, respectively, which were produced using random variables.
	(f) Comparison between the observed intensity profile (blue) and the simulation obtained by taking the square of FFT amplitude of the model in (e).
	(g) (0, 0, $L$) scans near $L = 6$ at various temperatures in zero field.
	(h) The integrated intensity of the main magnetic peaks at $q\sim0.4$ (red, $I_{\sim 0.4}$), $q\sim 0.5$ (blue, $I_{0.5}$), and remnant component (cyan, $I_{\text{ICM}}$) in the $L$-scan for (g) (see the inset).
	}
\end{figure}

We note that in the previous polycrystalline study \cite{EuZnGe} the magnetic susceptibility suggests successive magnetic transitions at 12.2 K and 9.5 K.
We also observed anomalous features in $M$-$T$ curves under the application of $\mu_{0}H=5$ mT, as shown in Fig. 3(c).
Double-peak feature in $M/H$ for $H\perp c$ at $T=10.5$ K and 9.0 K is similar to the previous results \cite{EuZnGe}.
By taking temperature derivative of $M$, we observe a peak at $T_{\text{N}}$ in $\text{d}M/\text{d}T$.
For $H\parallel c$, the transition near $T_{\text{N}}$ manifests as a dip and a broad peak near $T^{*}=11.3$ K implies another critical temperature while no obvious anomaly is observed in $C_{\text{p}}$ or $\Delta L/L$ (see dashed lines in Figs. 3(a)-(b)).
We observed distinct responses between in-plane and out-of-plane resistivities to these transition(-like) temperatures (Fig. 3(d)).
A sharp peak in $\rho_{zz}$ occurs only at $T_{\text{N}}$, which correlates with the onset of the out-of-plane modulation as observed in similar Eu-based helimagnets \cite{shang2021anomalous,riberolles2021magnetic,bauhofer1985electrical}.
$\rho_{xx}$, on the other hand, takes a broad peak at $T^{*}$ and exhibit a weak kink at $T_{\text{N}}$.
These transport anomalies imply that $T^{*}$ corresponds to the steep growth of in-plane magnetic correlations, which may originate from low-dimensionality of Eu triangular-lattice layers.

In order to clarify the magnetic structure of EuZnGe, we performed RXS.
We used the Eu $L_2$-edge resonance to observe the magnetic scattering due to Eu spin ordering.
The intensity profile in the (0, 0, $L$) scan is shown in Fig. 4(a).
We observed magnetic Bragg scattering around $L = 7.5-7.6$, which disappears at $T=$ 15 K (above $T_{\text{N}}$ and $T^*$).
We found a double-peak feature with a dominant peak at $L=7.6$ and a weak peak at $L=7.5$, which correspond to magnetic modulations along the $c^{*}$ axis with $q=\frac{2}{5}$ and $q=\frac{1}{2}$, respectively.
The energy profiles at (0, 0, 7.6) and (0, 0, 7.5) (see Fig. 4(b)) show a resonant peak near the Eu $L_2$ edge, $E\sim$ 7.615 keV, clarifying the Eu origin of the magnetic Bragg scattering.
We performed polarization analysis for the $q=\frac{2}{5}$ peak and confirmed the comparable intensities for $\pi'$ and $\sigma'$ channels, which is consistent with a spiral modulation.

We note that the double-peak feature as well as the diffusive intensity in intermediate $0.4<q<0.5$ cannot be reconciled with the proper-screw (helical) spin configuration with $q=0.4$, which corresponds to undistorted interlayer-rotation of 72$^{\circ}$ to exhibit a single peak at $L=7.6$.
Applying the broken-helix recently proposed in a related Eu-based intermetallics \ce{EuIn2As2} \cite{riberolles2021magnetic} is unlikely because in the present case the magnetic unit cell of $q=0.5$ (four-sublattice) is incompatible in that of $q=0.4$ (five-sublattice).
To reproduce the intensity profile, we introduce two types of spin configuration as shown in Figs. 4(c)-(d), and consider the actual spin structure as a random mixture of the two. 
Figure 4(c) shows a five-sublattice spin configuration for $q=0.4$ with a four-fold 60$^{\circ}$ inter-layer spin rotation followed by a 120$^{\circ}$ rotation (60$^{\circ}$-60$^{\circ}$-60$^{\circ}$-60$^{\circ}$-120$^{\circ}$).
Figure 4(d) corresponds to a four-sublattice model with 60$^{\circ}$-120$^{\circ}$-60$^{\circ}$-120$^{\circ}$ spin rotations.
Such a model with discretized spin rotations by 60$^{\circ}$ or 120$^{\circ}$ is hypothesized in the limit of strong hexagonal in-plane anisotropy, which is consistent with the site symmetry of Eu ions.
We construct a real-space spin configuration on a 1D-chain with $\sum _{i}n_i$ sites (see Fig. 4(e)), where $n_i$ is the number of sites in each local $i$-th section in which the spins modulate in either of the above two models.
We randomly distribute the configuration for $q=0.4$ or $q=0.5$ onto the $i$-th section by a ratio of 7:3, and determined the number $n_i$ by a random variable among $0\sim 50$.
Figure 4(f) compares the square of the fast Fourier transformation (FFT) of the spin configuration and the observed magnetic scattering intensity, where we obtain a reasonable agreement.
In particular, the present model reproduces the intensity in the intermediate $q$ region, which cannot be reconciled with a simple multi-domain model for $q=0.4$ and 0.5.

\begin{figure}[t]
	\includegraphics[width =  \columnwidth]{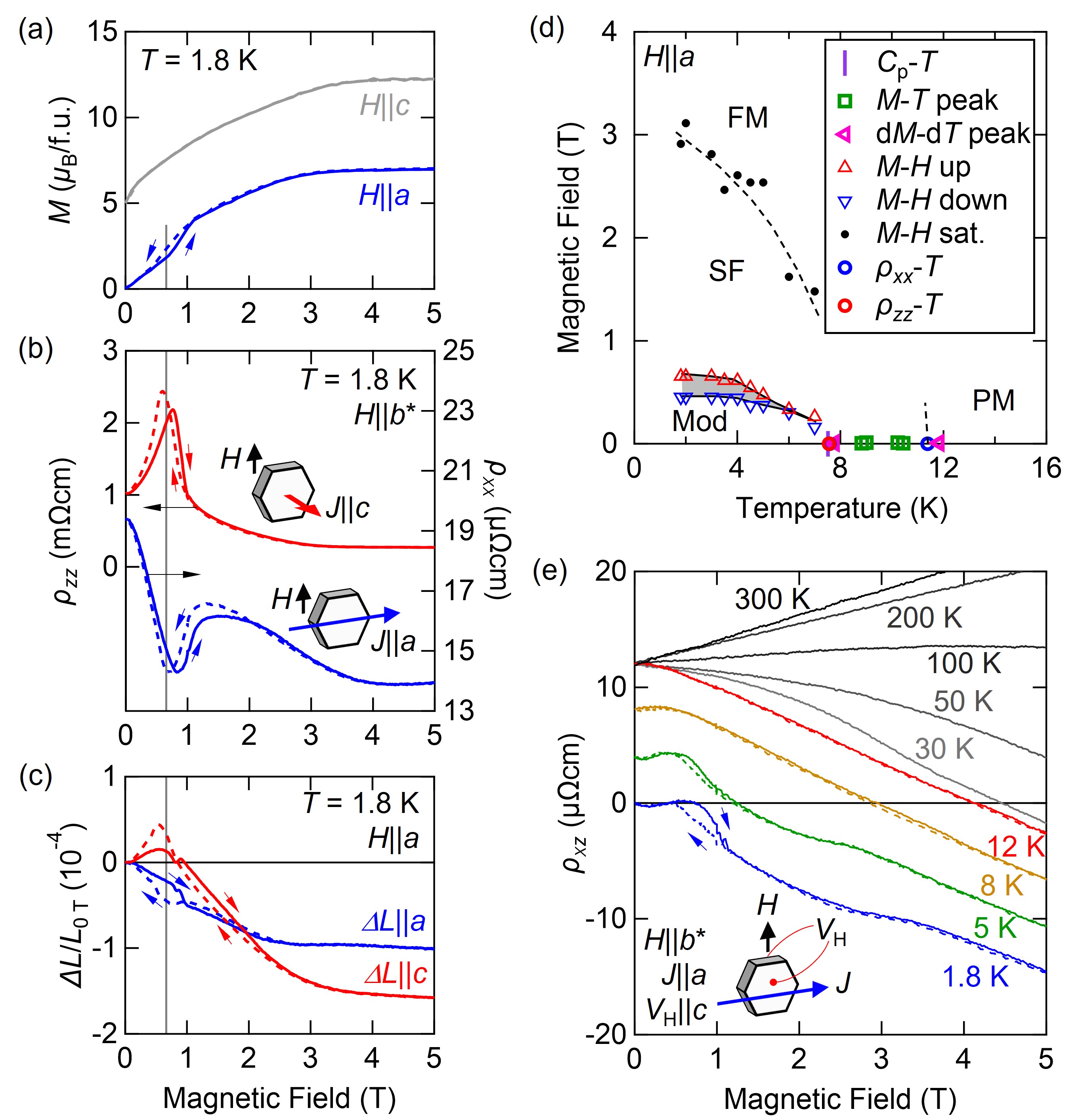}
	\caption{\label{fig5}(a)-(c) Field dependence of physical properties of EuZnGe for $H\parallel b^*$ at $T$ = 1.8 K.
	(a) magnetization ($M$), (b) out-of-plane resistivity ($\rho_{zz}$) and in-plane resistivity ($\rho_{xx}$), and (c) magnetostriction ($\Delta L_a/L_a$, $\Delta L_c/L_c$).
	Solid (dashed) curve is for field increase (decrease) process.
	For (a), $M$ under $H\parallel c$ with the offset of 5 $\mu_{\text{B}}$ is also shown by gray curve.
	(d) $H$-$T$ phase diagram for $H\parallel a$.
	Shaded area is hysteretic region.
	PM: paramagnetic; Mod: modulated; SF: spin-flopped; FM: field-induced ferromagnetic phases, respectively.
	(e) Hall resistivity ($\rho_{xz}$) at various temperatures. 
	Solid (dashed) curve is for field increase (decrease) process, and curves are shifted for clarity.
	}
\end{figure}

Figure 4(g) shows the intensity profile in the (0, 0, $L$) scan near $L=6$ at selected temperatures.
The intensity for $q(=L-6)=0.5$ ($I_{0.5}$) together with the intermediate diffusive intensity ($I_{\text{ICM}}$) disappear above $T=$ 8 K, while that of $q\sim$ 0.4 ($I_{\sim 0.4}$) persists with the shift to $q=0.38$.
We propose a realization of incommensurate smooth helical structure (namely the proper-screw) near $T_{\text{N}}$, likely due to thermal fluctuations.
As shown in Fig. 4(h), the double-peak structure disappears above $T_{\text{N}}$, which is correlated with the sharp peak in $\rho _{zz}$ (Fig. 3(d)).

\begin{figure}[t]
	\includegraphics[width =  \columnwidth]{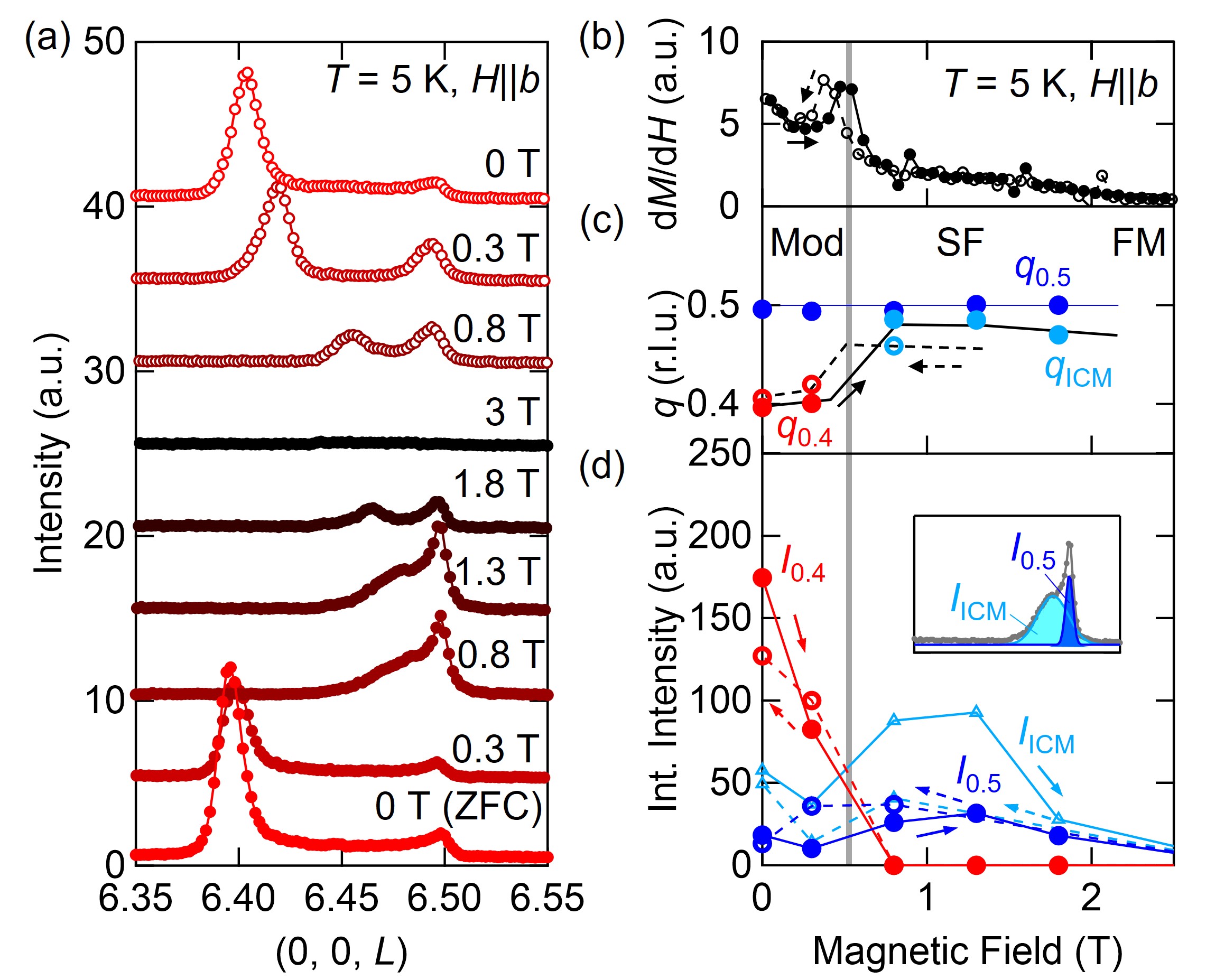}
	\caption{\label{fig6}(a) (0, 0, $L$) scan for magnetic RXS at various magnetic field for $H\parallel b$ at $T =$ 5 K.
	Prior to the measurement, the sample was cooled down in zero field. 
	(b)-(d) Field dependence of (b) $\text{d}M/\text{d}H$, (c) $q$ position of two main magnetic Bragg peaks (red: $q_{0.4}$, blue: $q_{0.5}$, cyan: $q_{\text{ICM}}$ (incommensurate $q$ in the SF phase)), and (d) the integrated intensity at $q_{0.4}$ (red, $I_{0.4}$), at $q_{0.5}$ (blue, $I_{0.5})$, and remnant intensity (cyan, $I_{\text{ICM}}$) (for Mod (SF) phase, see inset of Fig. 4(c) (Fig. 6(d))).
	Closed (open) symbol is for field increase (decrease) process, and solid (dashed) line in (c) is guide to the eyes.
	}
\end{figure}

Responses in a magnetic field clarify the coupling between magnetism and the underlying electric structure.
We measure magnetization for both $H\parallel c$ and $H\parallel a$ (Fig. 5(a)), and observe the saturation to near 7 $\mu_{\text{B}}$ for $H >$ 4 T, consistent with the divalent nature of Eu.
We note that the metamagnetic transition at $\sim$0.6 T \cite{EuZnGe} is reproduced only for the $H\perp c$.
It apparently corresponds to the transition from spiral to fan or transverse conical spin structure, as predicted in the magnetization process of helimagnets \cite{nagamiya1968helical}.
The resistivity shows distinct responses with respect to the direction of the electric currents (Fig. 5(b)).
$\rho_{zz}$ shows a sharp peak at the critical field, as is commonly observed in metamagnetic transitions \cite{mun2010physical,baranov1993effect}.
In the metamagnetic phase (SF phase), $\rho_{zz}$ shows a large negative magnetoresistance, implying the suppression of the out-of-plane modulation.
In contrast to this, $\rho_{xx}$ shows a dip, which may be attributed to enhanced conduction at domain walls between regions with different $q$.
We also observe magnetostriction associated with the metamagnetic transition (Fig. 5(c)).
In contrast to $\Delta L_a/L_a$, $\Delta L_c/L_c$ is nonmonotonic at the phase boundary, suggesting the coupling between spin configuration and inter-layer distance of triangular-lattices.
We place the anomalies observed in various physical properties measurements for $H\perp c$ in the $H$-$T$ phase diagram (Fig. 5(d)).
The peaks of $M-T$ curve for $H\perp c$ (green squares) slightly deviate from the transition temperatures estimated for $M-T$ in $H\parallel c$, and $\text{d}M/\text{d}T$, $C_{\text{p}}$, $\rho_{xx}$, and $\rho_{zz}$ in zero field.

The Hall resistivity also exhibits an intriguing field evolution with $H\parallel b^*$ (Fig. 5(e)).
At $T=$ 1.8 K, $\rho_{xz}$ shows a valley-like feature selectively in the SF phase compared with a monotonic negative slope at $T=$ 12 K, suggesting the dominant electron-type carrier transport at low temperatures.
It is reminiscent of the THE due to the emergence of noncoplanar spin configuration as proposed in the distorted-spiral compound \ce{YMn6Sn6} \cite{ghimire2020competing,wang2021field}.
Nevertheless in the present case, it is difficult to disentangle possible topological Hall response from field-modulation of the scattering time evidenced in $\rho_{xx}$ (Fig. 5(b)) as the Hall signals across the metamagnetic transition in intermetallics with high-mobility carriers closely resemble that of $\rho_{xx}$ through the multi-carrier transport \cite{ye2017electronic}.
In Fig. 5(e), $\rho_{xz}$ shows a sign change above 100 K, clearly indicating the multi-carrier nature.

To gain the insight on the field-induced SF phase, we measured the RXS profile under different magnetic fields.
Figure 6(a) shows that entering the SF phase at 0.8 T (see the peak in $\text{d}M/\text{d}H$ in Fig. 6(b) defining the phase boundary) induces an enhancement of the peak at $q=0.5$ and in contrast an abrupt shift of the modulation at $q=0.4$ to $q_{\text{ICM}}\sim 0.47$.
Interestingly, the double-peak structure with $q=0.5$ and $q_{\text{ICM}}$ survives up to $\mu_{0}H=$1.8 T and disappears in the FM phase at $\mu_{0}H=$ 3 T.
Through the removal of the applied $H$, the double-peak structure recovers from the SF phase (0.8 T), and eventually settles near identical profile  (but weaker intensity at $q=0.4$) in the zero field. 
The hysteretic behaviors of the $q$ position and the intensity of the magnetic scatterings are summarized in Figs. 6(c)-(d).
The step wise shift from $q=0.4$ to $q_{\text{ICM}}$ (Fig. 6(c)) coincides with the increase of the integrated intensity for $q=0.5$, $I_{0.5}$ (Fig. 6(d)).
The double $q$-peak structure in the SF phase excludes the simple fan or transverse conical spin configuration (described by a single-$q$) with no scalar spin chirality.
Further study is necessary to reveal possible noncoplanar spin configuration in the SF phase to understand the THE-like response in $\rho _{xz}$(Fig. 5(e)).

\section{Conclusion}
In conclusion, we succeeded in growing single crystals of stoichiometric magnetic semimetal EuZnGe and found magnetotransport properties strongly coupled with underlying spin structure.
Resonant x-ray scattering revealed the helical modulation along the $c$-axis, which is not a simple proper-screw, but randomized with spin-skipping in terms of the rotation in the hexagonal plane.
We found that the in-plane magnetic field induces the spin-flop-like metamagnetic transition.
The strong correlation with the longitudinal and transverse transport responses suggests that the present magnetic semimetal is potentially promising for emergent phenomena due to spin-conduction-electron coupling in spiral magnets \cite{yokouchi2020emergent,jiang2020electric}.

\begin{acknowledgements}
T.K. was financially supported by MEXT Leading Initiative for Excellent Young Researchers (JPMXS0320200135), JSPS KAKENHI Grant-in-Aid for Young Scientists B (No. 21K13784).
M.G. was supported by the JSPS KAKENHI Grant-in-Aid for Scientific Research (No. 20J10988).
S.K. was supported by JSPS KAKENHI Grant-in-Aid for Early-Career Scientists (No. 22K14010).
This work was partly supported by JSPS KAKENHI Grant-in-Aid for Scientific Research on Innovative Areas "Quantum Liquid Crystals" (No. JP19H05826 and No.19H01835).
The synchrotron radiation experiments were performed at SPring-8 with the approval of the Japan Synchrotron Radiation Institute (JASRI) (Proposal No. 2022A1751).
The resonant x-ray experiment at PF was performed under the approval of the Photon Factory Program Advisory Committee (Proposal No. 2020G665).
This work was partly performed using the facilities of the Materials Design and Characterization Laboratory in the Institute for Solid State Physics, the University of Tokyo.
T.K. acknowledges A. Kikkawa and R. Ishii for their advice and help to handle tantalum crucibles, and H. Nakao for the experiment at BL-3A, PF.
S.K. thanks K. Adachi and D. Hashizume for in-house x-ray diffraction characterization of the crystal quality.

\end{acknowledgements}

\appendix*
\section{Crystal structure of EuZnGe at 300 K}
Due to the small difference in the number of electrons between Zn (30) and Ge (32) it is difficult to distinguish two atoms by in-house XRD experiment.
Therefore, we performed synchrotron XRD experiments on BL02B1 at SPring-8 in Japan \cite{sugimoto2010extremely} using a high-quality single crystal of EuZnGe.
A two-dimensional detector CdTe PILATUS, which had a dynamic range of $\sim 10^7$, was used to record the diffraction pattern.
Diffraction data collection for crystal structural analysis was performed using a RIGAKU RAXIS IV diffractometer.
Intensities of equivalent reflections were averaged and the structural parameters were refined by using Jana2006 \cite{petvrivcek2014crystallographic}.

The results of the structural analysis of EuZnGe at 300 K are summarized in Tables I and II.
When the Zn and Ge sites are completely mixed, the length of the $c$-axis can be regarded as half, resulting in the absence of the $l=2n+1$ reflections.
We, however, found $l=2n+1$ reflections (Fig. 1(e)).
The degree of the site mixing at the Zn and Ge sites was further analyzed.
Although the occupancy of each of the Zn and Ge sites was refined by restricting the total value to 1, no signs of the site mixing were found, where the $R$ factors were not improved and each occupancy of the Zn and Ge sites remained almost unchanged (see Tables II).

\begin{table*}[b]
\centering
\caption{Structural parameters of EuZnGe at 300 K.
The space group is $P6_3/mmc$ (No. 194), and $a=b=$ 4.3700(5) \AA, $c=$8.5994(5) \AA, $\alpha = \beta = 90^{\circ}$, $\gamma =120^{\circ}$}

\begin{tabular}{l*{8}{c}}
\hline
\hline
 & Wyckoff & $x$ & $y$ & $z$ & $U_{11} (=U_{22})$ (\AA$^2$) & $U_{33}$ (\AA$^2$) & $U_{12}$ (\AA$^2$) \\
\hline
Eu & $2a$ & 0 & 0 & 0 & 0.00915(2) & 0.00967(3) & 0.004574(12) \\
\hline
Zn & $2d$ & 2/3 & 1/3 & 1/4 & 0.00580(3) & 0.01649(5) & 0.002901(13) \\
\hline
Ge & $2c$ & 1/3 & 2/3 & 1/4 & 0.00746(3) & 0.01649(6) & 0.003728(13) \\
\hline
\hline
\end{tabular}
\end{table*}

\begin{table*}[h]
\centering
\caption{Crystallographic data of EuZnGe}

\begin{tabular}{l*{2}{c}}
\hline
\hline
Temperature (K) & 300 \\
Wavelength (\AA) & 0.310109 \\
Crystal dimension ($\mu$m$^3$) & 50$\times$50$\times$50 \\
Space group & $P6_3/mmc$ \\
$a$ (\AA) & 4.3700(5) \\
$c$ (\AA) & 8.5994(5) \\
$Z$ & 2 \\
$F(000)$ & 250 \\
$(\sin \theta /\lambda)_{\text{max}}$ (\AA$^{-1}$) & 1.79 \\
$N_{\text{Total}}$ & 13080 \\
$N_{\text{Unique}}$ & 1281 \\
Average redundancy & 10.211 \\
Completeness (\%) & 99.96 \\
\hline
\multicolumn{2}{c}{$R$ factors when assuming the perfect ordering of Zn and Ge} \\
\hline
$N_{\text{parameters}}$ & 8 \\
$R_1$ ($I>3\sigma $) [number of reflections] & 1.23\% [868] \\
$R_1$ (all) [number of reflections] & 2.51\% [1281] \\
$wR_2$ (all) [number of reflections] & 2.95\% [1281] \\
GOF (all) [number of reflections] & 1.85\% [1281] \\
\hline
\multicolumn{2}{c}{$R$ factors when allowing the site mixing of Zn and Ge} \\
\hline
$N_{\text{parameters}}$ & 10 \\
$R_1$ ($I>3\sigma $) [number of reflections] & 1.23\% [868] \\
$R_1$ (all) [number of reflections] & 2.52\% [1281] \\
$wR_2$ (all) [number of reflections] & 3.77\% [1281] \\
GOF (all) [number of reflections] & 1.84\% [1281] \\
Occupancy of Zn/Ge at (2/3, 1/3, 1/4) & 1.034(8) / $-$0.034(8) \\
Occupancy of Ge/Zn at (1/3, 2/3, 1/4) & 0.994(7) / 0.006(7) \\
\hline
\hline
\end{tabular}
\end{table*}

\clearpage
\bibliography{reference}
\end{document}